\documentclass[%
 reprint,
 amsmath,amssymb,
 aps,
prd,
longbibliography]{revtex4-1}

\pdfoutput=1
\usepackage{graphicx}
\usepackage{dcolumn}
\usepackage{bm}
\usepackage[colorlinks=true,linkcolor=black,citecolor=blue,urlcolor=blue]{hyperref}

\newcommand{\mathsym}[1]{{}}
\def\id{\protect{{1 \kern-.28em{\rm l}}}}
\def\be{\begin{eqnarray}}
\def\ee{\end{eqnarray}}

\def\ha{\tfrac{1}{2}}

\def\a{\alpha}

\def\g{\gamma}

\def\det{\hbox{det}}

\def\Tr{{\rm Tr}}

\def\l {\lambda}

\def\O{{\mathcal O}}
\def\m{\mu}

\def\foot{\footnote}
\newcommand{\rf}[1]{(\ref{#1})}

\def\no{\nonumber}

\def\la{\label}
\def\l{\lambda}

\def\p{\phi}

\def\varpi{{\rm w}}

\def\del{\partial}
\def\s{\sigma}

\def\ed{\end{document}}
\newcommand{\mc}{\mathcal }
\def\iffa{\iffalse}

\def\d{\delta}

\def\Lie{\operatorname{Lie}}

\def\ddt{\frac{\ud}{\ud t}}

\def\ed{\end{document}}

\def\cg{c\,}

\def\g{\gamma}

\def \foot{\footnote}\def \l {\lambda}\def \iffa {\iffalse}
 \def \a  {\alpha} \def \ha {{1\ov 2}}
\def \ed {\end{document}}

\def \la {\label}

\def \ha {{1 \over 2}}

\def \cg {{\rm c}}

\def \ccg {{\cg_{_G}}}

\def\cG{\ccg}

\def \ha {\tfrac{1}{2}}

\newcommand{\arxivlink}[1]{\href{http://arxiv.org/abs/#1}{[arXiv:#1]}}
\usepackage{enumitem}
\usepackage{xcolor}

\newcommand{\ud}{\mathrm d}

\usepackage{bbm}

\hypersetup{breaklinks=true}

\begin{document}

\title{Equivalence of 1-loop RG flows in \texorpdfstring{\\}{}
4d Chern-Simons and integrable 2d sigma-models
} 

\vspace{-0.1cm}
\author{Nat Levine}\email{nat.levine@phys.ens.fr}

\affiliation{ \vspace{0.3cm}
Laboratoire de Physique {\rm and} Institut Philippe Meyer, \\ \'Ecole Normale Sup\'erieure, \\
  \mbox{  Universit{\'e} PSL, CNRS, Sorbonne Universit{\'e}, Universit{\'e} Paris Cit{\'e},} \\
24 rue Lhomond, F-75005 Paris, France\\
\vspace{-0.0cm}
}%

\date{\today}

\begin{abstract}
\vspace{-0.1cm}
We argue for the matching of 1-loop divergences between 4d Chern-Simons theory with Disorder defects and the corresponding integrable 2d $\sigma$-models of non-ultralocal type. Starting from the 4d path integral, we show under general assumptions that the 1-loop divergences localise to the 2d defects. They match the `universal' formulae developed in \arxivlink{2209.05502} for the 1-loop RG flows of integrable 2d $\sigma$-models. Our argument uses an alternate path integral representation for the 1-loop effective action and the known classical equivalence between the 4d and 2d theories.
\end{abstract}

\maketitle

\section*{Introduction}
Integrability in the context of 2d $\s$-models was first noted by Pohlmeyer, who discovered that the well-known sine-Gordon model, and generalisations of it, may be obtained by a certain `reduction' from	 $\s$-models \cite{Pohl}. More recently, in the context of string theory and AdS/CFT, such integrable $\s$-models have found useful application as exactly solvable worldsheet theories for strings in curved backgrounds, most notably in the case of AdS$_5 \times$S$^5$. In that instance of the AdS/CFT correspondence, the integrability on the string worldsheet  has been found to translate to a remarkable integrability for the scaling dimensions of the dual planar $\mc N=4$ super Yang-Mills theory (see the reviews \cite{AF,Beisert}).

Naturally, many authors have asked whether this setup can be modified while preserving the planar integrability: with less supersymmetry \cite{Pomoni}, in different dimensions \cite{Bab}, etc. On the string side, one may search for $\s$-models that (i) are integrable and (ii) have supergravity solutions as their target spaces.
 Another approach is to consider integrable deformations of a known setup, such as the string $\s$-model on AdS$_5 \times$S$^5$.
A zoo of various integrable deformations, meeting some or all of these criteria, have been found, including the $\beta$-deformations \cite{Lunin}, $\eta$-deformations \cite{Delduc:2013qra,Klimcik:2002zj}, and $\lambda$-deformations \cite{Hollowood:2014qma,Sfetsos:2013wia}. Overall, however, the general understanding of the space of integrable $\s$-models --- with supergravity target spaces or not --- remains challenging.

Recently, a new framework was proposed to systematise the landscape of integrable 2d theories: a  4d Chern-Simons theory that should act as a `parent' to many integrable 2d models \cite{CWY1,CWY2,CY3}.  
The two extra spacetime dimensions are parametrised by the complex spectral parameter of the integrable 2d theories.
The 4d  action, previously introduced in \cite{Costello1,Costello2}, is 
\begin{align}
&S_{4d} = \frac{i}{16\pi^2} \, \int  \phi(z) \, \ud z \wedge {\rm CS}(A) \ ,  \la{4d2}\\
&{\rm CS}(A) = \Tr[ A\wedge \ud A + \tfrac{2}{3}A \wedge A\wedge A] \ , \no \\
&  A = A_\m(x) \,  \ud x^\m   \in \Lie(G)  \ ,\qquad  x^\m=(z,\bar z\, ; \, \xi^+,\xi^-)\ . \no
\end{align}
Here the `twist function' $\phi(z)$ is a fixed meromorphic function, which, together with boundary conditions, will specify the 2d theory. We allow $\phi(z)$ to have both zeros and poles: the zeros are `Disorder' defects where the gauge field can become singular while leaving the action finite; the poles are 2d defects where appropriate boundary conditions must be imposed.

In the present context with Disorder defects, the correspondence between integrable 2d field theories and 4d Chern-Simons has been established only classically \foot{On the other hand, for the `Order' defects of \cite{CY3}, quantum aspects have been discussed in the talk of M.~Yamazaki at the \textit{Strings 2018} conference and in the recent paper \cite{Y}.
 Quantum effects were also considered in the context of integrable lattice models \cite{CWY1,CWY2,Y}.}. It was shown in a broad class of cases \cite{unif, Benini, Lacroix:2020flf} that
the 4d action localises to the defects, 
yielding integrable 2d $\s$-models of
 \textit{non-ultralocal} \foot{`Non-ultralocality' refers to the presence of derivatives of delta-functions in the classical Poisson brackets. This is a well-known obstruction to finding a direct integrable quantisation (see, e.g., \cite{NU2,NU3,NU4,NU1}).} 
 type in the $(\xi^+,\xi^-)$ directions \footnote{Our conventions are stated in Appendix \ref{conv}.}.

 It is therefore an important problem to clarify the extension of this correspondence to the quantum level. 
We will compute the 1-loop divergences in 4d Chern-Simons in a natural scheme, finding that they localise to the 2d defects and coincide with the 1-loop RG flow of the corresponding 2d field theory.

\section*{Review of classical equivalence \la{c4}}
The classical derivation of integrable 2d field theories from 4d Chern-Simons theory with Disorder defects was introduced in \cite{CY3} and elaborated in \cite{Vicedo,unif,Lacroix:2020flf,Benini}.

The 4d Chern-Simons action \rf{4d2} does not depend on the component $A_z$; equivalently, it has a gauge symmetry,
\be A_z \to A_z + \l_{ z} \ , \la{44} \ee which can always be used to fix $A_z=0$. It also has the standard 4d vector gauge invariance,\be A_\mu \to h^{-1} A_\mu h + h^{-1} \del_\m h\ ,\la{4gt} \ee where the gauge parameter $h\in G$ will eventually be constrained by the presence of the defects.

Varying the action \rf{4d2} with respect to $A_\mu$, we find  the following bulk equations of motion
\be \begin{aligned}
&\phi(z) F_{\bar z + }(A) = 0 \ , \\
&\phi(z)  F_{\bar z - } (A) = 0 \ , \\
  &\phi(z)  F_{+-}(A) = 0 \ . \la{eom}
\end{aligned} \ee
There is also a boundary term,
\be
\delta S \propto \int \ud^2 z \ \ud^2\xi \ \del_{\bar z} \phi(z)   \, \Tr[ A_+ \d A_- -  A_- \d A_+ ]  \  , \la{bt}
\ee	
which is localised to the set $P= \left\{\text{poles of }\phi(z)\ud z\right\}$ (the support of $\del_{\bar z} \phi(z) $). Hence, the poles of $\phi$ are naturally interpreted as 2d defects where one must fix boundary conditions to arrange the vanishing of \rf{bt}. A general class of admissible boundary conditions was constructed in \cite{unif,Lacroix:2020flf,Benini}.

For our purposes, the key assumptions are that the boundary conditions: (i) kill the boundary term \rf{bt} in the variational problem,
and (ii) satisfy a `Bianchi completeness' assumption specified below.
For simplicity, we may assume that the action \rf{4d2} converges at the poles of $\phi$: this means that either the poles are simple or the boundary conditions render the integral finite. The general case should work similarly, upon replacing the action \rf{4d2} with regularised one of \cite{Benini}.
 We also assume, as in \cite{unif}, approprate reality conditions on the group $G$ and the twist function $\phi$ so that the action is real; and that the 4d space is $\Sigma_2\times \mathbb{C}P^1$ where $\Sigma_2$ is a Euclidean or Lorentizan surface endowed with a flat metric.

The gauge field $A_\mu$ may be singular at the zeros of $\phi$; however, for a finite action, the singularity must be distributed among the components so that the  combination  $ \int \ud^2 z\, \phi(z)\,  \Tr\big[ A_{\bar z} [A_+ , A_-] \big]$ is finite. The zeros of $\phi$ are usually taken to be simple and partitioned into two equal sets, with $A_+$ blowing up at one set and $A_-$ at the other \cite{unif} --- but we will not need to use this property directly.

Writing $A_{\bar z} = - \del_{\bar z} g \,  g^{-1}$, 
the first two bulk equations in \rf{eom} become
\be
\phi(z) \, \del_{\bar z} L_\pm = 0 \ , \qquad L_\pm := g^{-1} A_\pm g  +g^{-1} \del_\pm g \ , 
\ee
i.e.\ $L_\pm$ is meromorphic in $z$ with poles only at the zeros of $\phi(z)$, and of the same order \foot{In \rf{ph} we give the expression for $L_\pm$ assuming $\infty$ is not a zero of $\phi(z)$, but its generalisation to include that case is very simple.}:
\be \begin{aligned}
&L_\pm = \mc A_\pm^{0} + \sum_{i=1}^k \sum_{p=1}^{m_i} \frac{\mc A_\pm^{i,p}}{(z-a_i)^p} \ , \\
&\phi(z) = R \frac{(z-a_1)^{m_1} \cdots (z-a_k)^{m_k}  }{(z-b_1)^{n_1} \cdots (z-b_l)^{n_l} } \ .\la{ph} 
\end{aligned} \ee
We denote by  $\mc A_\pm^{i,p}(\xi)$  the residues of the various poles; these are some 2d currents independent of $z$. 

Thus, on-shell, we may trade $A_\pm$ for a meromorphic 2d connection $L_\pm$. The remaining equation in \rf{eom} then imposes a zero-curvature equation for $L_\pm$,
\be
F_{+-}(L) \equiv \del_+ L_- - \del_- L_+ + [L_+, L_-] = 0 \ , \la{zc}
 \ee 
matching the equations of motion of a 2d integrable model with $L_\pm$ as its Lax connection.

\sloppy The 4d gauge invariance can be used to set $g=1$ `away' from the defects.
 At the defects, the boundary conditions considered in \cite{unif, Lacroix:2020flf,Benini} fix the form of the currents \mbox{$\mc A = \mc A(g|_{_P})$} in terms of $g$ (and its derivatives) at the poles $P$ of $\phi$. 
 Thus, after solving the first two equations in \rf{eom}, the only remaining degrees of freedom are $g|_{_P}$ and the classical 4d action localises to a 2d action at the defects,
\begin{align}
&S_{4d}(A)\to S_{2d}(g|_{_P}) \ . \la{rac}
\end{align}
The resulting 2d action generalises the known actions of many integrable 2d theories based on affine Gaudin models \cite{unif, Lacroix:2020flf,Benini,Yoshida,Stedman}.

\section*{Review of `universal' 1-loop divergences}\label{RU}
A new approach for the 1-loop renormalisation of integrable 2d $\s$-models was recently introduced by the present author \cite{Levine}. Firstly, we recall that the classical equations of motion of integrable 2d theories typically take a universal zero-curvature form \rf{zc} in terms of a Lax connection $ L_\pm (z, \mc A)$, which depends on the 2d fields through some currents $\mc A$, and meromorphically on the complex spectral parameter $z$. Secondly, 1-loop divergences are generally determined by the classical equations of motion. Therefore, it is natural for the 1-loop divergences of integrable theories to take a universal form in terms of the Lax connection.

An important subtlety 
is
that the zero-curvature equations \rf{zc} generally comprise not only on-shell equations of motion, but also some Bianchi identities satisfied off-shell. (These will eventually correspond to the boundary conditions in the 4d theory.)  It was explained \cite{Levine} by a general path-integral argument that this distinction between on-shell and off-shell equations is invisible at the level of 1-loop (on-shell) divergences. This argument in 2d made a  `Bianchi completeness' assumption
analogous to the one we will need to make for the boundary conditions in 4d.

The result can then be stated as follows. For different theories, the Lax connection can have different analytic structures, i.e.\ choices of poles in the $z$ plane.  For each analytic structure, the 1-loop divergences take a universal form in terms of the 2d currents multiplying the poles of the Lax. Remarkably, the divergences are independent of the particular form of the currents in terms of the fundamental 2d fields. 

For example, let us take a  Lax connection with only simple poles,
\be
L_+ = \sum_i \frac{1}{z-z_i} \mc A_+^i \ , \qquad L_- = \sum_j \frac{1}{z-w_j} \mc A_-^j \ , 
\ee
with all $z_i \neq w_j$. Then the divergent terms in the 1-loop effective action take the universal form \cite{Levine}
\be
\ddt \widehat S_{2d}^{(1)} = \frac{\cG}{2\pi} \sum_{i,j} \frac{1}{(z_i - w_j)^2}  \int \ud^2 \xi \ \Tr[ \mc A_+^i \mc A_-^j ]  \ ,  \la{sp}
\ee	
irregardless of the relation between $\mc A$ and the fundamental fields. Here, $\cG$ denotes the dual Coxeter number of the algebra ${\rm Lie}(G)$ in which the Lax is valued and $t=\log \mu$ where $\mu$ is the renormalisation scale.

For a given analytic structure of the Lax connection, universal formulas like \rf{sp} are easily computed from the following path-integral representation for the 1-loop effective action \cite{Levine} 
\be\begin{aligned}
&\widehat S_{2d}^{(1)}\big( \bar{\mc A}\big)  = -i \log \int \mc D a_\pm \  \mc D u^I \  \\
&\qquad \qquad \qquad \qquad \exp{i \int d^2 \xi \ u^I \, \text{zc}_I(\bar{\mc A} +a )} \ ,\quad \quad  \la{exp}
\end{aligned}\ee	
where the 2d vector $a_\pm$ and scalars $u^I$ are valued in $\Lie(G)$. We define $\{\text{zc}_I(\mc A) = 0\}_I$ as the finite set of 2d equations following from the zero-curvature equation \rf{zc}:
\be
F_{+-}(z,\mc A) = 0 \quad \forall z \quad \iff \quad {\rm zc}_I(\mc A) = 0 \quad \forall I \ . \ \
\ee

 In other words, the 1-loop effective action \rf{exp} is the `volume' of the space of flat connections with the prescribed choice of poles in the spectral variable. In light of the link to 4d Chern-Simons theory, it is tempting to rewrite this volume as a 4d path-integral:
\begin{align}
&\widehat S_{2d}^{(1)}\big( \bar{\mc A}\big)  \sim - i \log  \int [\mc D \ell_\pm]_{\text{fix poles}} \, \mc D u \la{expr} \\
&\qquad\qquad \qquad  \qquad \exp{i \int \ud^2\xi  \,  \ud^2 z \ \Tr\Big[ u \, F_{+-}\big( \bar L + \ell \big)\Big]}  . \no
\end{align}
Here the integral is over fluctuations $\ell_\pm$ of a Lax connection with fixed meromorphic dependence on $z$. 
In what follows, we shall 
derive a formula like \rf{expr}, and hence the universal 2d result \rf{exp}, 
from 4d Chern-Simons theory.

\section*{Matching 1-loop RG flows}
 Let us use  a background field expansion $A_m = \bar A_m + a_m$ to expand the 4d action around a classical solution (here $m=(\bar z, +, -)$),
\begin{equation}\begin{split}
\hspace{-0.1cm}S_{4d}[\bar A + a] = S_{4d}[\bar A] + \int \Tr[ a^m  \, \mc O_{mn}(\bar A) \, a^n ]+ \ldots \ . \la{lint}
\end{split}\end{equation} 
Here $\mathcal O_{mn}(\bar A) \,  a^n = 0$ are the linearised equations of motion and the integration symbol denotes integration against the 4-form $dz \wedge d\bar z \wedge d\xi^+\wedge d\xi^-$. Then the 1-loop effective action is given by the determinant of $\mc O$,
\begin{align}
\widehat S^{(1)}_{4d}[\bar A] &= -i \log \int_{\rm B.C.s} \frac{\mc D  a_m}{\rm gauge}  \, e^{i\,  \int \Tr[ a^m  \, \mc O_{mn}(\bar A) \, a^n] }  \\
&= \tfrac{i}{2} \log \det' {\O(\bar A)} \ . \la{det}
\end{align}
Here `B.C.s' denotes the imposition of the boundary conditions and $\det '$ means the determinant is taken over gauge-invariant states. Since the 4d action does not depend on $A_z$, we do not integrate over it.

\subsection*{1. Changing variables}
To evaluate the 1-loop determinant \rf{det}, we will use an alternate path-integral representation:
\begin{widetext}
\begin{align}
\widehat{S}_{4d}^{(1)} &= - \tfrac{i}{2} \log \int_{\rm B.C.s} \frac{\mc D  a_m \, \mc D  u^m}{\rm gauge}  \, \exp{i \int \, \Tr[u^m \,  \O_{mn}(\bar A) \,  a^n ] } \la{pint} \\
&= - \tfrac{i}{2} \log \int_{\rm B.C.s} \frac{ \mc D a_m \, \mc Du^m}{\rm gauge}  \, \la{p2}
  \exp{i \int \, \phi(z) \, \Tr[u^{\bar z} \, F_{+-}(A) + u^- \, F_{\bar z +}(A) + u^+ \, F_{\bar z -}(A) } ]\ , 
\end{align}
\end{widetext}
where we have input the 4d Chern-Simons equations of motion \rf{eom} and are implicitly expanding $A_m=\bar A_m+a_m$ to linear order.  This is the same trick that was used to derive the `universal formulae' \rf{sp},\rf{exp} in 2d \cite{Levine}: to this 1-loop order, the equations of motion are effectively imposed as Lagrange multiplier constraints. The  Lagrange multipliers $u^m$ naturally take the same boundary conditions as the physical fluctuations $a_m$.

 Following the classical discussion above, it is convenient to change variables as $(A_{\bar z}, A_\pm) \to (g, L_\pm)$ with
 \be \begin{aligned}
 &A_{\bar z} = - \del_{\bar z} g \, g^{-1} \ , \quad A_\pm = g L_\pm g^{-1} -\del_\pm g \,  g^{-1}  . \la{fr}
\end{aligned} \ee
We note that this introduces an additional gauge freedom,
\be\begin{aligned}
&g \to g \, q \, , \quad L_\pm \to q^{-1} L_\pm q + q^{-1} \del_\pm q \ , \\
& q\in G \, , \qquad \del_{\bar z} q = 0 \ , \la{egt}
\end{aligned} \ee
to which we will return later.

Instead of integrating over the fluctuations $a$ of $A$, we may equivalently integrate over the fluctuations $(\gamma,\ell_\pm)$ of $(g,L_\pm)$, defined as
\be
g = \bar g\, \g \ , \qquad L_\pm = \bar L_\pm + \ell_\pm \ .
\ee
  Then, rotating the Lagrange multipliers $u^m \to g \, u^m\,  g^{-1}$, the path integral \rf{p2} becomes
\begin{widetext}
\begin{align}
 \widehat{S}^{(1)}_{4d}
&= - \tfrac{i}{2} \log \int_{\rm B.C.s} \frac{\mc D  \g \ \mc D \ell_\pm \ \mc D  u^m}{\rm gauge}  \, \exp{i \int \, \phi(z)\,  \Tr[u^{\bar z} F_{+-}(L) + u^- \, \del_{\bar z} L_+ + u^+ \, \la{r2} \del_{\bar z} L_- ] } \ .
\end{align}
\end{widetext}

\subsection*{2. Solving meromorphicity equations}
First, let us integrate over  $u^\pm$ in \rf{r2}, imposing the equations $\phi(z) \, \del_{\bar z}  L_\pm=0$ as delta-functions. These equations force $L_\pm$ to take the form \rf{ph}:  meromorphic with poles only at the zeros of $\phi(z)$. The coefficients  $\tilde{\mc A}_\pm^{i,p}(\xi)$ of the poles are 2d fields (indicated hereafter by a tilde). We may thus trade the integral over $\ell_\pm$ for one over the fluctuations $\tilde \a_\pm^{i,p}$ of $\tilde{\mc A}_\pm^{i,p}(\xi)$,
\begin{align}
&\int \mc D  \ell_\pm  \, \mc D u^\pm \, \exp{ i \int \phi(z) \, \Tr[ u^-  \, \del_{\bar z} L_+ + u^+ \, \del_{\bar z} L_-]}  \la{mm} \\
&\quad  = \int \mc D  \ell_\pm  \,\delta^{(4)}(\phi(z) \del_{\bar z} L_+)   \,\delta^{(4)}(\phi(z) \del_{\bar z} L_-)  =   \int  \mathcal D \tilde \a_\pm^{i,p}  \ , \no
\end{align}
where
\be
\tilde{\mc A}^{i,p} = {\bar{\mc A}}^{i,p} + \tilde \a^{i,p}  \ , \ \  L_\pm =\tilde{\mc A}_\pm^{0} + \sum_{i=1}^k \sum_{p=1}^{m_i} \frac{\tilde{\mc A}_\pm^{i,p}}{(z-a_i)^p}  \ .\ \ \ \ \ \
\ee
The identity \rf{mm} is correct up to quantum corrections, related to the Jacobian for the transformation
\be \begin{aligned}
&L_\pm \to \big(\phi(z) \, \del_{\bar z} L_\pm\,  , \ \mc A_\pm^{i,p}(L)\big) \ ,\la{traj} \\
&\mc A_\pm^{i,p}(L) := L_\pm (z-a_i)^{p_i}\big|_{z=a_i} \ .
\end{aligned} \ee
This Jacobian is expected to be a power-law divergent constant (independent of fields). We shall neglect such constant terms in the effective action, corresponding only to an overall normalisation of the path integral. While there is no standard dimensional regularisation scheme removing power divergences for this non-Lorentz-invariant 4d theory, these constant terms may eventually be interpreted as defining the appropriate measure in 4d.
 We will return to this issue in the \textit{Discussion} section.

As a result, the divergent (non-constant) terms in the effective action \rf{r2} are given by
\begin{align}
\widehat{S}^{(1)}_{4d} 
&\sim - \tfrac{i}{2} \log \int_{\rm B.C.s} \frac{\mc D  \g  \ \mc D  \tilde \a^{i,p}_\pm \ \mc D u^{\bar z}}{\rm gauge}\la{r3} \\
&\qquad \qquad\exp{i \int \, \phi(z)\,  \Tr[u^{\bar z} F_{+-}(L(z,\tilde{\mc A} ) )] } \ . \no
\end{align}

\subsection*{3. Gauge fixing}
While $g$ (or its fluctuation $\g$) does not appear in the action in \rf{r3}, it generally still appears in the boundary conditions, denoted `B.C.s'.
The B.C.s are assumed to impose a particular form of the Lax currents,
\be 
\tilde{\mc A} =\tilde{\mc A}(\tilde g|_{_P}) \ ,  \la{BCs}
\ee
in terms of some `defect'  components $\tilde g|_{_P}$ of the field $g$  (certain combinations of $g$ and its $z$-derivatives evaluated at $P$). We will split the integral over the fluctuation $\g$ as 
\be
\mc D \g = \mc D \tilde \g|_{P} \  \mc D \g|_{\rm bulk} \ , \la{fact}
\ee
where 
$\g|_{\rm bulk}$ denotes the rest of the field $\g$, constrained to match $\tilde \g|_{_P}$ at $P$. We are not being careful about possible constant measure factors in \rf{fact} --- see \textit{Discussion} section and Appendix \ref{Mf}.

The standard 4d gauge transformation \rf{4gt} acts as $g \to h^{-1} g $ but must be constrained to preserve the boundary conditions. A sufficient choice is to insist the gauge transformation acts trivially on the defect components $\tilde g|_{_P}$ (this is the choice made in \cite{Lacroix:2020flf} that successfully reproduces the classical actions of many integrable 2d models). Thus $g|_{\rm bulk}$ can be fixed to any field configuration interpolating smoothly between $\tilde g|_{_P}$ (there are many). The result is to integrate over only $\tilde g|_{_P}$ (or rather its fluctuation $\tilde \g|_{_P}$),
\begin{align}
\widehat{S}^{(1)}_{4d} 
&\sim - \tfrac{i}{2} \log \int_{\rm B.C.s} \frac{\mc D  \tilde \g|_{_P}  \ \mc D  \tilde \a^{i,p}_\pm \ \mc D u^{\bar z}}{\rm gauge}\la{r3a} \\
&\qquad \qquad\exp{i \int \, \phi(z)\,  \Tr[u^{\bar z} F_{+-}(L(z,\tilde{\mc A} ) )] } \ . \no
\end{align}

An additional gauge invariance \rf{egt} was introduced by the change of variables \rf{fr}. This invariance can
 be fixed on some component of $\tilde g|_{_P}$ (e.g. the value of $g$ at one of the defects). Conventionally \cite{unif,Benini, Lacroix:2020flf}, $\phi(z) \, \ud z$ is assumed to have a double pole at $z=\infty$ with boundary condition $A_\pm|_{\infty}=0$; then one may fix $\tilde g|_{\infty}=1$. 
If one were instead to keep this gauge symmetry, it would become the standard 2d gauge symmetry of the Lax connection's zero-curvature equations.

\subsection*{4. Boundary conditions}
Upon integrating over $\tilde \g|_{_P}$ in \rf{r3a}, the boundary conditions \rf{BCs} will impose some Bianchi identities on $\mc A$,
\begin{align}
\hspace{-0.35cm}\int_{\rm B.C.s} \mc D \alpha_\pm^{i,p} \ &\mc D \tilde\g|_{_P} =  \int \mc D \alpha^{i,p} \ \mc D \tilde\g|_{_P} \,  \delta \big(\tilde{\mc A} - \tilde{\mc A}(\tilde g|_{_P})\big) \la{ep}\\
&= \int \mc D \alpha_\pm^{i,p}\,  \prod_s  \delta( B_s \cdot \alpha )\\
&=\int \mc D \alpha_\pm^{i,p}\, \mc D \tilde v^s  \, \exp{i \int d^2 \xi \  \tilde v^s \, (B_s \cdot \tilde \alpha) } \, . \la{ed}
\end{align}
Here the linearised Bianchi identities $B_s \cdot \tilde \alpha=0$ are indexed by $s$ and are being imposed by 2d  Lagrange multipliers $\tilde v^s$.

\bigskip
\noindent\textit{{Bianchi Completeness Assumption:}}\ \  Let us assume that the linearised Bianchi identities $B_s \cdot \alpha=0$ following from the boundary conditions are a subset of the linearised zero-curvature equations $F_{+-}\big(L(z,\bar{\mc A}  + \tilde {\a}) \big)=0$.

\bigskip
This assumption is apparently met by all of the boundary conditions considered in \cite{unif}; we conjecture that it is related to the 2d theory being a $\s$-model. It is analogous to the assumption made when deriving the `universal' 2d RG flow \cite{Levine} that we are matching.

Since the zero-curvature equations are already being imposed by Lagrange multipliers in \rf{r3a}, then the exponential factor in \rf{ed} can be discarded and the effective action is given by
\begin{align}
& \widehat{S}^{(1)}_{4d}
\sim - \tfrac{i}{2} \log \int\frac{ \mc D\tilde v^s \ \mc D \tilde \a_\pm^{i,n} \ \mc D u^{\bar z}  }{\rm gauge}\la{3e}\\
&\qquad \qquad  \qquad \quad \exp{i \int \, \phi(z)\,  \Tr[u^{\bar z} F_{+-}(L(z,\tilde{\mc A} ) )] } \ .\no
\end{align}
The `universality' of the 2d RG flow appears here as the disappearance of the boundary conditions. 

Discarding a constant measure factor $\int \frac{\mathcal D \tilde v^s}{\rm gauge}$ (see Appendix \ref{Ogs}), one obtains 
\begin{align}
& \widehat{S}^{(1)}_{4d}
\sim - \tfrac{i}{2} \log \int\frac{\mc D \tilde \a_\pm^{i,n} \ \mc D u^{\bar z}  }{\rm gauge}\la{3f}\\
&\qquad \qquad  \qquad \quad \exp{i \int \, \phi(z)\,  \Tr[u^{\bar z} F_{+-}(L(z,\tilde{\mc A} ) )] } \ .\no
\end{align}
This is equivalent to the universal 2d result \rf{exp}: it is the `volume' of solutions to the same set of equations, now written in 4d language  --- and precisely matches the prediction \rf{expr} from 2d. To see their formal equivalence, note that \rf{3f} still has a gauge invariance, as only certain modes of $u^{\bar z}$ appear in the action. Gauge fixing all other modes to zero (see Appendix \ref{Ogs} for careful discussion) the 4d path integral \rf{3f} simply reduces to the universal 2d one \rf{exp},
\be
\widehat{S}^{(1)}_{4d} \sim \widehat{S}^{(1)}_{2d} \ ,
\ee
where $\sim$ denotes equivalence of divergent non-constant terms.

\section*{Discussion}
We have shown a formal identity between the 1-loop divergences of 4d Chern-Simons theory with Disorder defects and the corresponding integrable 2d $\s$-models, at the level of the effective action. We assumed 
 a particular type of boundary conditions satisfying a `Bianchi completeness' property. 

We neglected field-independent measure factors, i.e.~the overall normalisation of the path integral.
 Such normalisations depend on the choice of regularisation and we may expect to absorb them into the definition of the 4d path integral measure. 
In any case, even if one regards the measure as an extra divergence in the 4d theory compared to the 2d one, 
 the link between the 4d and 2d RG flows is only minimally modified and statements such as renormalisability unaffected.

We believe this to be the first quantum check of the 4d Chern-Simons proposal in the presence of Disorder defects. 
In principle,   since $\phi(z)$ 
 appears in the combination $\phi(z)/\hbar$, one should worry that its zeros (cf.\ the $\hbar\to \infty$ limit) may lead to a breakdown of the semi-classical expansion \cite{CWY1}.
 However, we found that at least the 1-loop divergences are well-behaved. 
This could, of course, begin to fail at higher-loop orders.

Another possibility is that, even at higher loops, the 4d Chern-Simons formulation is the `correct' setup, consistent with integrability, for quantising integrable $\s$-models. The relation to the 2d models may then be subject to quantum corrections, yielding the $\a'$ (i.e.\ $\hbar$) corrections that are understood to be ubiquitous in the target spaces of  integrable $\s$-models \cite{HLT1,HLT2,LT,BW,BVW,Alfimov}. 

A useful application of the present work would be to replace the standard 2d computation of the 1-loop RG flow with a direct 4d computation. 
This would be helpful since, at present, the 2d universal formula \rf{exp} still requires a 1-loop computation for each choice of poles of the Lax connection. On the other hand, a 4d computation might address all pole structures (i.e.\ all twist functions) at once. 
In future, we hope to attempt such a 4d computation and to establish equivalence  of the universal formula \rf{exp} with the RG flows posited elsewhere in the literature \cite{Delduc:2020vxy,Derryberry:2021rne,CTA}.

\begin{acknowledgments}
I am grateful to K.~Costello, N.~Ishtiaque, S.~Lacroix, A.~A.~Tseytlin, B.~Vicedo and A.~Wallberg  for useful discussions and/or comments on this manuscript, and to S.~Lacroix and A.~Wallberg for a related collaboration. This work was supported by the Institut Philippe Meyer at the \'{E}cole Normale Sup\'{e}rieure in Paris.
\end{acknowledgments}

\appendix

\section{Conventions \la{conv}}
We use the 2d co-ordinates $(\xi^+,\xi^-)$, which may be regarded as 2d light-cone co-ordinates $\xi^\pm = \ha(\tau \pm \sigma)$ in Lorentzian signature or as a single complex co-ordinate $(\xi^+,\xi^-) = (w, \bar w)$ in Euclidean signature. 

We denote by $z$ the complex spectral parameter of the 2d theory, which constitutes the 3rd and 4th dimensions in the 4d theory.

The 4d co-ordinates are denoted together as \mbox{$x^\mu = (z,\bar z; \, \xi^+, \xi^-)$}, with $\mu = (z,\bar z, +, -)$. We also use a restricted index $m=(\bar z, +, -)$ omitting the $z$ direction.\sloppy

The 2d Lax connection $L_\pm$ and the 4d gauge field $A_\mu$ are valued in the Lie algebra $\Lie(G)$ of some Lie group $G$, which is taken to be semisimple.

Classical actions are denoted as $S$ and 1-loop effective actions as $\widehat S^{(1)}$.

In the \textit{Matching RG flows} section, we indicate all 2d fields with a tilde to distinguish them from 4d fields.

\section{Principal Chiral Model example \la{Pex}}
In this appendix, we illustrate the argument above for the simple example of the Principal Chiral Model (PCM). To build it from 4d Chern-Simons, one should start with the twist function
\be
\phi(z) = h\, \frac{(1-z)(1+z)}{z^2}  \ . 
\ee
We designate that $A_+$ may blow up at $z=-1$, and $A_-$ at $z=1$.  To engineer the PCM, we should choose the boundary condition $A_\pm|_{0,\infty}=0$ at the double poles $0$ and $\infty$ of $\phi(z) \ud z$ \cite{unif} --- one can check that these conditions meet the criteria set out above. 

Writing down the path integral \rf{p2} for the 1-loop divergences, changing variables $A \to(g, L)$ as in \rf{fr} and solving the meromorphicity equations $\phi(z) \, \del_{\bar z} L_\pm = 0$, we find that the Lax connection takes the meromorphic form
\be
L_\pm = \tilde{\mc A}_\pm^0  + \frac{1}{1\pm z}\,\tilde{ \mc A}_\pm  \ . \la{LA}
\ee
Following the steps above with this Lax connection, the 1-loop effective action may be written as the path integral \rf{r3},
\begin{align}
\widehat{S}^{(1)}_{4d} \
&\sim - \tfrac{i}{2} \log \int_{\rm B.C.s} \frac{\mc D  \g  \  \mc D  \tilde \a_\pm^0 \ \mc D \tilde \a_{\pm} \  \mc D u^{\bar z}}{\rm gauge}  \la{rr3}\\
&\qquad \exp{i \int \, \phi(z)\,  \Tr\Big[u^{\bar z} \big(F_{+-}(\tilde{\mc A}^0) + \tfrac{1}{1-z^2}\, D^0 \cdot \tilde{\mc A} }\no\\
&\qquad \qquad \qquad \qquad \qquad \qquad +  \tfrac{z}{1-z^2}\, F_{+-}^0(\tilde{\mc A}) \big)\Big] \no \ , \\
&\hspace{-0.3cm}D^0 := \del + {\rm ad}_{\tilde{\mc A}^0} \ , \no \\
&\hspace{-0.3cm}F^0_{+-}(\tilde{\mc A}) := D^0_+ \tilde{\mc A}_- - D^0_- \tilde{\mc A}_+ + [\tilde{\mc A}_+, \tilde{\mc A}_-]   \ . \no
\end{align}
We remind the reader that the tilde denotes that $\tilde{\mc A}$ are 2d fields, and we use the split $\tilde{\mc A} = \bar{\mc A} + \tilde{\a}$ into a background field and a fluctuation. 

The boundary conditions impose that
\be
\tilde{\mc A}^0_\pm = \tilde g_\infty^{-1} \del_\pm \tilde g_\infty \ , \quad \tilde{\mc A}_\pm = \tilde g_0^{-1} \del_\pm \tilde  g_0 -  \tilde g_\infty^{-1} \del_\pm \tilde g_\infty\ .  \ \ \  \la{bP}
\ee
where $\tilde g_0, \tilde g_\infty\in G$ are the values of $g$ at the defects.  We split the integral over fluctuations $\g$ of $g$ to the values at the defects $0$ and $\infty$, and the values elsewhere,
\be
\mathcal D \g = \mathcal D \tilde \g_0 \ \mathcal D \tilde \g_\infty \  \mathcal D \g|_{\rm bulk} \ .
\ee

The 4d gauge symmetry is completely fixed on $\g|_{\rm bulk}$, leaving only the values $\tilde g_0$ and $\tilde g_\infty$ at the defects. Then fixing the additional gauge freedom \rf{egt} as $\tilde g|_\infty=1$,  the boundary conditions \rf{bP} become
\be
\tilde{\mc A}^0_\pm = 0 \ , \quad \tilde{\mc A}_\pm = \tilde g_0^{-1} \del_\pm \tilde  g_0 \ .  \la{bcg}
\ee
Upon integrating over $\tilde g_0$ in \rf{rr3}, the boundary condition \rf{bcg} imposes a Bianchi identity $F_{+-}(\tilde{\mc A})=0$:
\begin{align}
\widehat{S}^{(1)}_{4d} \
&\sim - \tfrac{i}{2} \log \int\frac{ \mc D \tilde \a_{\pm} \, \mc D u^{\bar z}}{\rm gauge} \ \delta\big(F_{+-}(\tilde{\mc A})\big) \la{rr4}\\
& \exp{i \int \, \phi(z)\,  \Tr\Big[u^{\bar z} \big(\tfrac{1}{1-z^2}\, \del \cdot \tilde{\mc A}  +  \tfrac{z}{1-z^2}\, F_{+-}(\tilde{\mc A}) \big)\Big]} \no \ .
\end{align}
The Bianchi Completeness Assumption is indeed satisfied since the Bianchi identity is implied by the zero-curvature equations. Therefore, after stripping off a constant measure factor (this is just the result of overcounting in \rf{p2} equations of motion that are trivialised by the boundary conditions --- see Appendix \ref{Ogs}), we obtain
\begin{align}
\widehat{S}^{(1)}_{4d} \
&\sim - \tfrac{i}{2} \log \int\frac{ \mc D \tilde \a_{\pm} \, \mc D u^{\bar z}}{\rm gauge} \la{rr5}\\
& \exp{i \int \, \phi(z)\,  \Tr\Big[u^{\bar z} \big(\tfrac{1}{1-z^2}\, \del \cdot \tilde{\mc A}  +  \tfrac{z}{1-z^2}\, F_{+-}(\tilde{\mc A}) \big)\Big]} \no \ .
\end{align}
Gauge fixing all but two modes of $u^{\bar z}$ to zero as discussed in Appendix \ref{Ogs}, this 4d path integral reduces to a 2d one imposing the same equations:
\begin{align}
\widehat{S}^{(1)}_{4d}
&\sim - \tfrac{i}{2} \log \int  {\mc D \tilde \a_{\pm} \, \mc D \tilde V \, \mc D \tilde W}  \la{PR}\\
&\qquad \exp{i \int \ud^2 \xi \, \Tr\Big[\tilde V  \, \big(\del \cdot \tilde{\mc A} \big)  +  \tilde W \,  F_{+-}(\tilde{\mc A}) \Big] }\no \ .
\end{align}
The 2d Lagrange multipliers $\tilde V,\tilde W \in \Lie(G)$ are the remaining modes of $u^{\bar z}$.

Eq.\ \rf{PR} is a special case of the 2d universal formula \rf{exp}, and may be computed by a single 1-loop diagram, yielding a particular case of the result \rf{sp} above \cite{Levine},
\be
\ddt S^{(1)}[\bar{\mc A}] = - \frac{\cG}{8\pi}  \, \int \ud^2 \xi \ \Tr[\bar{\mc A}_+ \bar{\mc A}_-] \ . \la{d1}
\ee
Under the map \rf{rac} between classical actions, the 4d action with this twist function and boundary conditions becomes the PCM action,
\be
 S [\bar{\mc A}] = -\frac{ h}{8\pi} \,  \int \ud^2 \xi \  \Tr[\bar{\mc A}_+ \bar{\mc A}_-] \ . \la{d2}
\ee
Hence the divergent 1-loop term \rf{d1} signals an RG flow of the coupling $h$ in \rf{d2} given by
\be
\ddt h = \cG \ ,
\ee
which is indeed the well-known 1-loop beta-function of the PCM.

\section{Other gauge symmetries \la{Ogs}}
To rewrite \rf{3e} as \rf{3f}, we stripped off a constant measure factor $\int \frac{\mathcal D \tilde v^s}{\rm gauge}$. More properly, this factor may be attributed to a gauge symmetry,
\be
\tilde v^s  \to \tilde v^s + \tilde \lambda^s \ . \la{vg}
\ee
originating from the boundary conditions killing some degrees of freedom that were initially overcounted in \rf{p2}. 
As a result, the Lagrange multipliers  $\tilde v^s$ are completely decoupled in \rf{3e}. Gauge fixing $\tilde v^s=0$ then yields eq.~\rf{3f}.

Now let us explain why the 4d path integral \rf{3f} is equivalent to the 2d `universal' result \rf{exp}. The exponent in \rf{3f} only depends on a few `modes' of the Lagrange multiplier $u^{\bar z}$ in the $z$ directions. Explicitly, supposing (with $I$ a finite index)
\be
F_{+-}\big( L( z, \tilde{\mc A})\big) = \sum_I f^I(z)  \, {\rm zc}_I(\tilde{\mc A}) \ ,
\ee
then only the modes $\tilde{u}^I := \int d^2 z \, \phi(z) \, u^{\bar z} \, f^I$  of the field $u^{\bar z}$ appear in \rf{3f}, and it is invariant under the local 4d transformation (constrained by boundary-type conditions)
\be
u^{\bar z} \to u^{\bar z} + \lambda^{\bar z} \ , \qquad \int \ud^2 z \ \p(z) \  \l^{\bar z} \, f^I = 0 \ . \la{guf}
\ee
Picking some `bi-orthogonal' functions $g_I(z,\bar z)$ satisfying $\int d^2z \, \p(z) \ f^I \, g_J = \d^I{}_J$, then one may fix the gauge
\be
u^{\bar z} = \sum_I \Big(\int d^2 z \ \phi(z) \ u^{\bar z} \, f^I \Big)\  g_I  = \sum_I \tilde{u}^I  \, g_I\ . \la{ug}
\ee
Finally, after gauge fixing, one should only integrate over this finite set of modes $\tilde{u}^I(\xi)$, and \rf{3f} becomes precisely the `universal'  2d result \rf{exp}.

\bigskip
Our prescription has been to treat all local symmetries of the linearised theory \rf{lint} as gauge symmetries in 4d: let us explain why this is natural. This is immediate for the standard 4d gauge symmetry \rf{4gt}. While one may generally consider different choices of boundary conditions for this gauge transformation at the defect --- i.e.\ different choices of gaugings at the defect --- the present choice (trivial gauge action on $g|_{_P}$) is the standard one that was used to obtain the classical actions of many integrable 2d theories \cite{Lacroix:2020flf}. It is also natural to treat shifts \rf{44} of $A_z$ as a gauge symmetry, since the 4d action simply does not depend on $A_z$.

Meanwhile, the redundancy \rf{egt} is an artefact our change of variables \rf{fr}, so should certainly be treated as a gauge symmetry, despite only depending on three of the co-ordinates. The two gauge symmetries discussed above in this appendix are
  artefacts of our overcounting of the equations of motion in the path integral \rf{p2} to compute the 1-loop determinant \rf{det}. We listed the linearised equations of motion, each being imposed by a Lagrange multiplier. After solving two bulk equations of motion, the third one localised to equations on the 2d defects;  however, we naively overcounted it as a 4d equation, leading to the redundancy \rf{guf}. Additionally, some of the degrees of freedom on the 2d defects were removed by the boundary conditions, but we had naively counted these equations, leading to the redundancy \rf{vg}.

\section{Measure factors \la{Mf}}
As explained in the \textit{Discussion} section, we have not been careful about the overall normalisation of the path integral, i.e.\ a possibly divergent constant term in the effective action. On the other hand, we have been careful about any field-dependent divergences. In this appendix, we will clarify why certain steps in the above argument 
do not generate field-dependent divergences.

First, let us consider the choice of path integral measure for the fluctuation $\g$ of the group-valued field $g$ in \rf{fact}. In splitting the path integral into defect and bulk parts, there could in principle be an additional measure factor, which may be divergent. Still, if the path integral is to be invariant under group rotations, it is natural that such a factor should not depend on the background field~$\bar g$.

For the Jacobian of the transformation \rf{traj}, it is enough to note that the transformation is linear in fields.

Next, we turn to the manipulations of the path integral from eq.\ \rf{ep} to \rf{ed}. Since those path integrals are purely 2d ones, it makes sense to define the path integral in the standard way for relativistic 2d theories (it is only in the 4d theory that there is not relativistic invariance). We may use dimensional regularisation, in which power law divergences are absent and we only need to deal with log-divergences. We then note that the Jacobians appearing in these manipulations will not produce log-divergences, since they are determinants of first-order \textit{chiral} operators, so any log-divergence would not be Lorentz-invariant.

Finally, let us explain why the Faddeev-Popov determinants for all gauge fixings here are trivial. For the gauge fixings in the main text, and for \rf{vg}, this is obvious because the gauge fixing conditions are of the form $X - \text{const} = 0$, where the gauge transformation is $\delta X = \lambda$. For the gauge symmetry \rf{guf}, the gauge fixing condition \rf{ug} is
\be
G:=u^{\bar z} -\sum_I \Big(\int d^2 z \ \phi(z) \ u^{\bar z} \, f^I \Big)\  g_I  = 0 
\ee
Since the gauge transformation acting as $\delta u^{\bar z} = \lambda^{\bar z}$ is constrained by
\be
\int \ud^2 z \ \p(z) \  \l^{\bar z} \, f^I  = 0 \ , 
\ee
then the gauge variation of $G$ is trivial: $\tfrac{\delta G}{\delta \lambda^{\bar z}} = \tfrac{\delta u^{\bar z}}{\delta \lambda^{\bar z}} = \mathbbm{1}$.

\bibliography{ChernNote}

\end{document}